\documentclass[universe,review,submit,oneauthor,pdftex]{Definitions/mdpi} 

\firstpage{1} 
\pubvolume{7}
\issuenum{10}
\articlenumber{360}
\pubyear{2021}
\copyrightyear{2021}
\externaleditor{Academic Editor: Olga Mena and
Pablo Fern\'andez de Salas} 
\datereceived{3 September 2021} 
\dateaccepted{22 September 2021} 
\datepublished{27 September 2021} 
\hreflink{https://doi.org/10.3390/
universe7100360} 
\pdfoutput=1

\usepackage[inter-unit-product =\cdot]{siunitx}

\Title{Status of Anomalies and Sterile Neutrino Searches at Nuclear~Reactors}

\TitleCitation{Status of Anomalies and Sterile Neutrino Searches at Nuclear Reactors}

\Author{Stefan Schoppmann 
 $^{1,2}$ \orcidA{}}

\AuthorNames{Stefan Schoppmann}

\AuthorCitation{Schoppmann, S.}

\address{%
$^{1}$ \quad Department of Physics, University of California, Berkeley, CA 94720-7300, USA; stefan.schoppmann@berkeley.edu\\
$^{2}$ \quad Lawrence Berkeley National Laboratory, Berkeley, CA 94720-8153, USA}

\abstract{Two anomalies at nuclear reactors, one related to the absolute antineutrino flux, one related to the antineutrino spectral shape, have drawn special attention to the field of reactor neutrino physics during the past decade. Numerous experimental efforts have been launched to investigate the reliability of flux models and to explore whether sterile neutrino oscillations are at the base of the experimental findings. This review aims to provide an overview on the status of experimental searches at reactors for sterile neutrino oscillations and measurements of the antineutrino spectral shape in mid-2021. The individual experimental approaches and results are reviewed. Moreover, global and joint oscillation and spectral shape analyses are discussed. Many experiments allow setting of constraints on sterile oscillation parameters, but cannot yet cover the entire relevant parameter space. Others find evidence in favour of certain parameter space regions. In contrast, findings on the spectral shape appear to give an overall consistent picture across experiments and allow narrowing down of contributions of certain isotopes.}



\begin{document}
\section{Introduction}
\label{sec:introduction}
During the last 20 years, measurements of neutrino oscillations in the three-flavour framework have determined all mixing angles and the magnitudes of the neutrino mass splittings~\cite{globalNeutrino1,globalNeutrino2,globalNeutrino3}.
However, beginning with the last generation of reactor experiments since 2010, two anomalies, one related to the absolute flux and one related to the spectral shape of reactor antineutrinos, emerged~\cite{5MeV1}.

The Reactor Antineutrino Anomaly (RAA) was discovered in a re-evaluation of the prediction of antineutrino spectra emitted by nuclear reactor cores in 2011~\cite{thesisMueller,conversion1,reactorSpectrum1}.
The re-evaluation was based on the so-called ab~initio method, which builds the total antineutrino spectrum from the sum of all available beta-branches of all fission products predicted by a nuclear simulation code. For~the most relevant $^{235}\text{U}$ and $^{239}\text{Pu}$ isotopes, updated information from nuclear databases and reference electron spectra of $^{235}\text{U}$, $^{239}\text{Pu}$, and~$^{241}\text{Pu}$ fissions were used~\cite{illFluxMeasurement1,illFluxMeasurement2}.
The re-evaluation led to an average normalisation shift, while maintaining the shape and uncertainty of the spectra similar to the previous analysis.
The new prediction revealed a \SI{6.5}{\percent} deficit between detected and predicted neutrino fluxes at baselines less than \SI{100}{\metre}. Depending on the prediction model~\cite{Berryman:2019hme}, the~significance of the RAA is up to 2.8 standard deviations~\cite{Gariazzo:2017fdh}.

Two hypotheses are mostly discussed in this context.
One is given by beyond-standard-model physics when assuming the existence of a fourth non-standard neutrino state leading to neutrino oscillations at short distances.
This state is expected to be sterile, i.e.,~non-weakly interacting.
If the mass splitting with respect to the known three active neutrino mass-eigenstates is around \SI{1}{eV}, significant oscillations towards this new neutrino eigenstate may be visible for reactor antineutrinos at baselines smaller than \SI{100}{\metre}.
As a result, a~deficit in the detected absolute antineutrino flux is observable at those baselines.
Such a sterile neutrino could also explain the lower than predicted electron neutrino rates measured in the calibration runs of the gallium solar neutrino experiments~\cite{Kaether:2010ag, Abdurashitov:2009tn}.
With a significance of 2.7~standard deviations, the~flux deficit triggered a new set of reactor antineutrino experiments at very short baselines of about 10\,m.
They are searching for characteristic oscillation patterns in connection to a sterile neutrino~state.

An alternative hypothesis for the explanation of the deficit addresses the antineutrino flux model itself.
The uncertainties of the summation method are sizeable. Moreover, the~predicted shape depends on the utilised databases. As~argued in~\cite{Hayes2014,Hayes:2015yka} a revision would allow a better understanding of the impact.
The alternative approach to the derivation of the flux model relies on single measurements of beta-spectra of uranium and plutonium irradiated by thermal neutrons~\cite{illFluxMeasurement1,illFluxMeasurement2,Hahn:1989zr,Haag:2013raa}.
From those spectra, the~conversion to antineutrino spectra, which yields the predicted antineutrino flux, is non-trivial~\cite{conversion1,conversion2}.
Erroneous steps in this method or an underestimation of its systematic uncertainties could also explain the RAA or reduce its significance~\cite{5MeV1,5MeV2}.
An updated calculation, which includes forbidden decays via nuclear shell model calculations leads to an increase in the flux at energies above 4 MeV~\cite{Hayen:2019eop}.
While this reduces the observed shape anomaly, it also yields an increase in the overall flux prediction and thereby an increased rate deficit.
In contrast, a~recent re-evaluation of reactor antineutrino spectra based on the measured ratio between cumulative $\beta$-spectra from $^{235}\text{U}$ and $^{239}\text{Pu}$ suggests resolving the rate deficit~\cite{Kopeikin:2021ugh}.
Aside from nuclear physics, it was also argued that common uncertainties in detector calibrations could contribute to the observed spectral shape anomaly~\cite{mention2017}.

\section{Searches for Sterile Neutrinos at~Reactors}
\label{sec:sterile}
Searches for sterile neutrinos at reactors can be carried out by two largely independent approaches.
On the one hand, the~absolute rate of antineutrinos can be compared to its prediction (rate approach).
On the other hand, distortions in the antineutrino spectra dependent on energy and distance from the antineutrino source can be investigated (shape approach).
While the rate approach requires good knowledge on the predicted flux, the~shape approach can, in~principle, be carried out in a relative fashion between data points at various energies and distances, largely independent from a prediction.
In both cases, the~detectable effects would be caused by the disappearance of electron antineutrinos via oscillations into sterile neutrinos.
Below 100\,m, the~survival probability of an initial electron antineutrino is given in good approximation in natural units as
\begin{equation}
 P_{\bar{\nu}_e \rightarrow \bar{\nu}_e} (L,E) = 1 - \sin^2(2 \theta_{ee}) \sin^2\left(\Delta m_{41}^2 \frac{L}{4 E}\right)
\end{equation}
where $L$ and $E$ are the baseline and antineutrino energy, respectively, $\theta_{ee}$ is the mixing angle, and~$\Delta m_{41}^2 = m_{4}^2 - m_{1}^2$ is the difference of the squared eigenvalues of the new mass eigenstate $m_{4}$ and the first mass eigenstate $m_{1}$.
The original RAA yields a sterile neutrino oscillation with $\left[ \sin^2(2\theta_{ee}) = 0.17, \Delta m_{41}^2 = \SI{2.3}{{eV}^2} \right]$~\cite{raa2012}.

In all experiments, nuclear reactors produce electron antineutrinos, which are detected by an inverse beta decay reaction (IBD):
\begin{equation}
\bar{\nu}_{e} + p^{+} \longrightarrow e^{+} + n .
\end{equation}

The coincidence signal of the prompt positron and the delayed neutron is a characteristic signature, which allows discrimination of many background events.
The neutron typically thermalises before its capture by a nucleus can take place.
Both the~thermalisation and the time constant of the capture process lead to the characteristic delay in the coincidence.
If a unloaded scintillator is used, the~capture happens via hydrogen.
Here, a~2.2\,MeV gamma is released after a mean capture time of about \SI{200}{\micro \second}.
To increase the probability of a neutron capture and for better background discrimination, the~scintillator can be doped with gadolinium (Gd).
In this case, the~capture energy is increased to about 8\,MeV.
Due to the higher cross-section for thermal neutron capture, the~coincidence time is reduced to typically \SI{30}{\micro \second} if the concentration of Gd is at the 0.1\,wt.\%-level.
Apart from Gd, $^{6}\text{Li}$ can be used which allows a better event localisation since the lithium decays into triton and an alpha particle after neutron~capture.

The IBD positron annihilates promptly, depositing its kinetic energy and two times 511\,keV from the annihilation.
Due to the higher mass of the neutron $m_n$ compared to the proton $m_p$, the~IBD reaction has a kinematic threshold on the neutrino energy of 1.8\,MeV.
The prompt energy deposition $E$ as seen by the detector is connected to the incident antineutrino energy $E_{\bar{\nu}_{e}}$ by
\begin{equation}
    E \sim E_{\bar{\nu}_{e}} - (m_n - m_p) + m_e
\end{equation}
where a small kinetic energy transfer to the neutron is~neglected.

As sources of the electron antineutrinos, nuclear reactors are particularly suited since they produce an intense, continuous, and~pure antineutrino flux.
The antineutrinos are produced in the $\beta$-decays of neutron-rich fission fragments in the reactor core.
A flux of more than $10^{20}$ antineutrinos per GW thermal power and second can be reached.
Two types of reactors are used in sterile neutrino searches: power reactors with lowly enriched $^{235}\text{U}$ (LEU) and compact research reactors with highly enriched $^{235}\text{U}$ (HEU).
While LEU reactors offer a much higher flux and therefore better statistics, HEU reactors offer a much more compact antineutrino source and thus a sharper oscillation baseline. Moreover, HEU reactors show less evolution of the fission fraction between fissile elements during a reactor cycle and offer more reactor-off times, where pure background data can be acquired by~experiments.

\subsection{Shape~Approach}
The shape approach in the search for sterile neutrinos has been carried out by several experiments at nuclear reactors during the last years.
The experimental designs vary between movable detectors that can cover an extended baseline-range and fixed setups that cover a smaller range.
Most detectors are segmented, to~allow simultaneous measurements at multiple baselines.
A high energy resolution, precise energy scale, and~a baseline around 10\,m are required for all experimental setups to test the RAA.
Highly segmented or moveable detectors allow the performance of the oscillation analysis in a relative fashion, thereby being largely independent of precise reactor flux predictions.
This especially includes any shape uncertainties due to the spectral shape~anomaly.

The collaborations employ different statistical techniques to derive exclusion contours or to infer allowed parameter spaces~\cite{Agostini:2019jup}.
Most notably, there are different choices in the method to derive confidence sets in the two-dimensional parameter space.
Those methods encompass a two-dimensional method (2D), a~one-dimensional raster-scan method (RS), and~a $\text{CL}_\text{s}$ method.
While these methods are in themselves correct, their use partly impairs a direct comparison of results and requires careful global or joint analyses to combine results~\cite{STEREO:2019ztb}.
Moreover, it has been pointed out that some of the experimental results are based on statistical assumptions which are prone to overestimation of the significance of discovery~\cite{StereoPROSPECT:2020raz,Coloma:2020ajw}.
This criticism especially addresses the use of Wilks' theorem in cases where its assumptions are not strictly met~\cite{wilks1938}.
Once again, this strengthens the importance of carefulness when combining results~globally.

\begin{figure}[H]
    \includegraphics[width=0.7\linewidth]{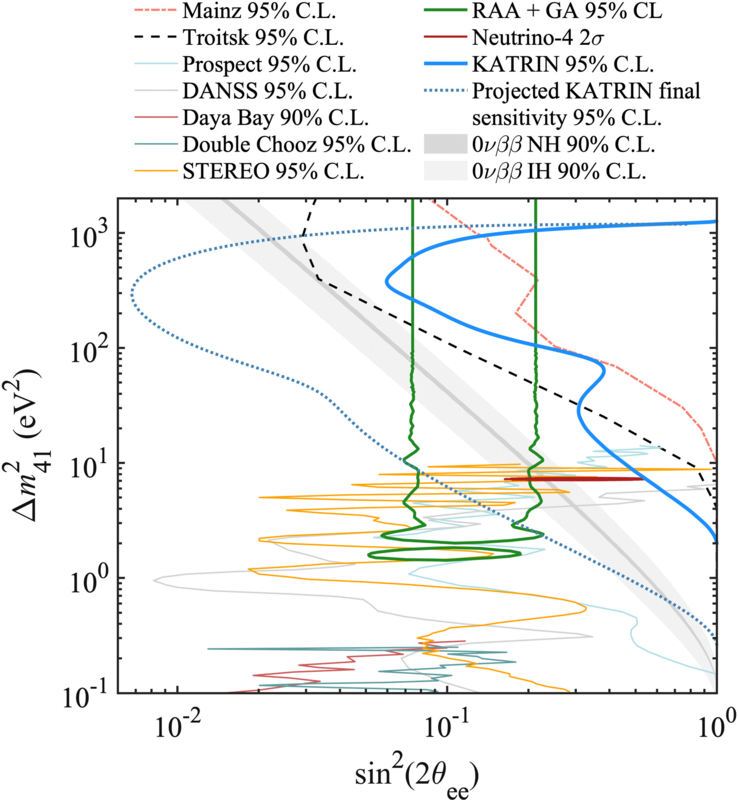}
    \caption{Exclusion contours of all reactor experiments in the plane of $\left[ \sin^2(2\theta_{ee}), \Delta m_{41}^2\right]$ alongside the allowed contours of the RAA and Gallium anomaly as well as Neutrino-4. KATRIN's current and expected exclusion limits are shown in addition. Reprinted from~\cite{KATRIN:2020dpx} under CC BY~4.0.}
    \label{fig:oscall}
\end{figure}

All reactor experiments and their results are discussed in the following and summarised in Figure~\ref{fig:oscall} and Table~\ref{tab:oscall}.
\begin{specialtable}[H]
    \caption{Comparison of the experimental parameters (reactor thermal power $P_\text{th}$, water equivalent overburden $D$, baseline $L$, target mass $m$), detection technique and statistical method used in the search for sterile neutrinos at reactors. If~more than one detector site or method exists, a~full list is~given.\label{tab:oscall}}
\setlength{\tabcolsep}{2.6mm}
    \begin{tabular}{lrrrrll}
        \toprule
    \textbf{Experiment} & \boldmath{$P_\text{th}$}\textbf{/MW} & \boldmath{$D$}\textbf{/m.w.e.} & \textbf{\boldmath{$L$}/m }& \textbf{\boldmath{$m$}/t} & \textbf{Detection} & \textbf{Method} \\
    \midrule
    SoLid & 80 & 10 & 6--9 & 1.6 & $^{6}\text{Li-PS}$ & n.a. \tabularnewline
    NEOS & 2800 & 20 & 24 & 1.0 & Gd-LS & RS \tabularnewline
    DANSS & 3100 & 50 & 11--13 & 0.9 & Gd-LS & $\text{CL}_\text{s}$ \tabularnewline
    STEREO & 58 & 15 & 9--11 & 1.7 & Gd-LS & RS, 2D, $\text{CL}_\text{s}$ \tabularnewline
    PROSPECT & 85 & 1 & 7--9 & 4.0 & $^{6}\text{Li-PS}$ & 2D, $\text{CL}_\text{s}$ \tabularnewline
    Neutrino-4 & 100 & 5--10 & 6--12 & 1.5 & Gd-LS & 2D \tabularnewline
    Daya Bay & 17,400 & 250, 860 & 550, 1650 & 80, 80 & Gd-LS & 2D, $\text{CL}_\text{s}$ \tabularnewline
    D-Chooz & 8500 & 120, 300 & 400, 1050 & 8, 8 & Gd-LS & RS \tabularnewline
    RENO & 16,800 & 120, 450 & 294, 1383 & 16, 16 & Gd-LS & 2D, $\text{CL}_\text{s}$ \tabularnewline
    \bottomrule
    \end{tabular}
\end{specialtable}
\unskip
\unskip

\subsubsection{SoLid}
The SoLid experiment~\cite{SoLid:2020cen} is located at baselines between 6 and 9\,m from the BR2 HEU
reactor in Belgium.
Its core has a height of 90\,cm and a diameter of 50\,cm, providing 80\,MW thermal power.
The detector, placed at an overburden of 10\,m.w.e., is foreseen to exploit a novel technique aiming for a high vertex resolution of the IBD event and a good neutron–gamma discrimination.
The detector target has a composite scintillator design made from $16 \times 16$ cubes of 5\,cm width consisting of polyvinyl toluene (PVT) scintillator, which are optically separated by reflective Tyvek.
Two faces of each PVT cube are covered by a layer of LiF:ZnS(Ag) to detect neutrons.
The signal to background ratio in SoLid is estimated to be 0.33.
No oscillation results have been reported by SoLid to~date.

\subsubsection{NEOS}
Among the first experiments, NEOS presented results in 2017~\cite{resultNEOS}.
NEOS is installed 24~m away from the LEU Hanbit nuclear power plant in Korea.
It has a thermal power of 2.8~GW and a core size of 3.1~m diameter and 3.8\,m height.
The detector is unsegmented and has a volume of $1\,\text{m}^{3}$ filled with a 0.5\% Gd-loaded liquid scintillator (Gd-LS).
The minimum overburden is 20\,m water equivalent (m.w.e.) and its signal to background ratio is 22.
NEOS compares its spectrum to Daya Bay's unfolded spectrum~\cite{resultDAYABAY} and excludes the parameter space below $\sin^{2}(2\theta_{ee})=0.1$ for values of $\Delta m^{2}_{41}$ between 0.2 and 2.3\,$\text{eV}^{2}$ at 90\% confidence level (CL).

\subsubsection{DANSS}
The DANSS experiment~\cite{resultDANSS,Danilov:2020ucs} has been operated since 2016 at a LEU reactor of 3.1\,GW thermal power in Kalinin, Russia, with a core of 3.7\,m height and 3.2\,m diameter.
The detector has variable baselines at 10.7\,m, 11.7\,m and 12.7\,m, is highly segmented by 2500 optically separated strips and consists in total of $1\,\text{m}^{3}$ plastic scintillator.
The polystyrene-based separators contain 6\% of Gd oxide.
The site has an overburden corresponding to a 50 m.w.e.~and gives a signal to background ratio of 20.
The oscillation analysis in DANSS excludes oscillations for a mass splitting between 0.5 and $2.5\,\text{eV}^{2}$ for mixing angles $\sin^{2}(2\theta_{ee})$ between 0.1 and 0.01 at 95\% CL.

\subsubsection{STEREO}
The STEREO experiment~\cite{STEREO:2019ztb} is located at the 58\,MW HEU reactor of the Institut Laue-Langevin (ILL), France, and~has collected data since 2016.
It measures neutrinos at baselines between 9 and 11\,m in its six optically separated cells and has a maximal overburden of 15 m.w.e.
The cells are filled with 1800 litres 0.2\% Gd-LS and surrounded by 2100 litres of Gd-free scintillator to convert escaping gammas.
The signal to background level is rather low at 1, due to nearby neutron beam lines.
STEREO excludes a parameter space between 0.8 and $4\,\text{eV}^{2}$ for mixing angles above $\sin^{2}(2\theta_{ee})=0.1$ at 95\% CL.

\subsubsection{PROSPECT}
PROSPECT~\cite{PROSPECT:2020sxr} uses an 85\,MW HEU reactor at the Oak Ridge National Laboratory in the USA with a core diameter of 0.4\,m and height of 0.5\,m at a baseline of 7 to 9\,m.
The detector is highly segmented into 154 strips using 3000 litres LS loaded with 0.1\% $^{6}\text{Li}$.
PROSPECT has an overburden of 1 m.w.e.
By using a pulse shape discrimination (PSD) technique ($^{6}\text{Li-PS}$), a~signal to background ratio of better
than 1 is possible.
A region between 0.8 and $4\,\text{eV}^{2}$ in the sterile oscillation
parameter space can be excluded at 95\% CL above $\sin^{2}(2\theta_{ee})=0.09$.

\subsubsection{Neutrino-4}
The Neutrino-4 experiment~\cite{Serebrov:2020kmd} started collecting data in 2016 at the SM-3 HEU reactor in Dimitrovgrad, Russia, which offers a compact core of $35 \times 42 \times 42 \,\text{cm}^{3}$ and exhibits 100\,MW thermal power.
Neutrinos are detected inside a 0.1\% Gd-LS detector segmented into  $5\times 10$ cells with an overall volume of $1.8\,\text{m}^{3}$.
The detector is moveable and offers baselines between 6 and 12\,m.
The signal to background ratio is 0.5.
In contrast to the above experiments, Neutrino-4 reports an oscillation signal with 2.9 standard deviations significance at $\sin^{2}(2\theta_{ee})=0.36\pm0.12\,\text{[stat]}$ and $\Delta m^{2}_{41}=(7.3\pm1.17)\,\text{eV}^{2}$.
The result is, however, partly excluded by the above results and in strong tension with constraints from cosmology~\cite{cosmologicalConstraints,sterileReview}.
The analysis of Neutrino-4 has been criticised regarding various \mbox{points~\cite{Danilov:2020rax,StereoPROSPECT:2020raz,Coloma:2020ajw},} most significantly regarding the soundness of the statistical~technique.

\subsubsection{Kilometre-Baseline~Experiments}
\begin{figure}[H]
    \includegraphics[width=0.9\linewidth]{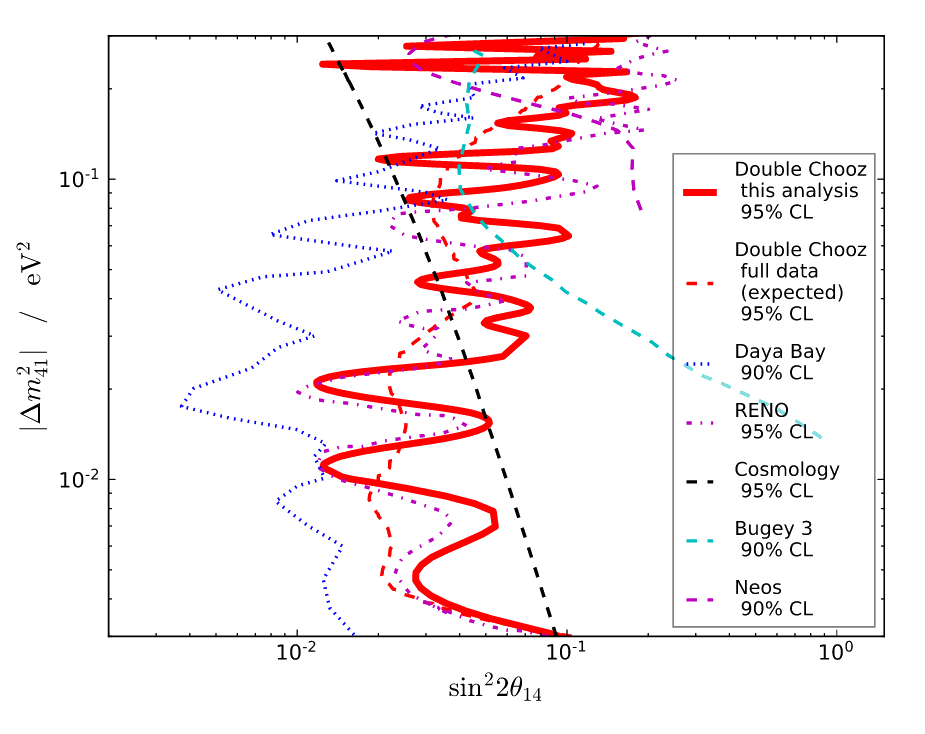}
    \caption{Exclusion contours of the kilometre-baseline reactor experiments in the plane of $\left[ \sin^2(2\theta_{ee}), \Delta m_{41}^2 \right]$ alongside NEOS' contour. Reprinted from~\cite{DoubleChooz:2020pnv} under CC BY~4.0.}
    \label{fig:kilometrebaseline}
\end{figure}

Apart from the above very short baseline experiments, the~previous generation of kilometre-baseline experiments, designed to measure $\theta_{13}$, have some sensitivity to sterile neutrinos at smaller mass splittings.
All three collaborations (Daya Bay, Double Chooz, RENO) use similar experimental setups with identical far detectors at a 1\,km baseline and near detectors at a few hundred metres.
The ton-scale detectors have unsegmented fiducial volumes and are filled with about a 0.1\% Gd-loaded scintillator.
Sterile oscillations could manifest in relative differences between the baselines or in deviations from the predictions.
Daya Bay was able to cover the region of $2\times 10^{-4}\,\text{eV}^{2} \leq m^{2}_{41} \leq 0.3\,\text{eV}^{2}$ finding no evidence of sterile neutrinos~\cite{DayaBay:2016qvc}.
Covering the region of $5\times 10^{-3}\,\text{eV}^{2} \leq m^{2}_{41} \leq 0.3\,\text{eV}^{2}$ Double Chooz also does not find an indication for sterile neutrinos~\cite{DoubleChooz:2020pnv}.
Similarly, RENO reports no evidence for sterile neutrinos covering a range of
$10^{-4}\,\text{eV}^{2} \leq m^{2}_{41} \leq 0.5\,\text{eV}^{2}$~\cite{RENO:2020uip}.
The results are summarised in Figure~\ref{fig:kilometrebaseline}.
\unskip

\subsubsection{Global and Joint~Analyses}
Since most experiments have published their main results or are close to releasing them, joint analyses are moving more into focus.
While global analyses take advantage of supplemental data released by some collaborations alongside their results, joint analyses between collaborations can allow a promising approach to exploit the full systematics between setups.
The latest global analyses find evidence for the existence of sterile neutrino oscillations with 2 to 3 standard deviations significance~\cite{globalSterileNeutrino1,globalSterileNeutrino2,sterileReview,Berryman:2020agd}.
However, results are partly in tension among each other or with cosmological constraints~\cite{cosmologicalConstraintsOnSterile}, dependent on the reactor flux model and uncertainties, and~vary with respect to the included datasets.
As some collaborations are updating their analyses and even extending their datasets, further changes in global results are expected in the future.
A first joint oscillation analysis was undertaken by NEOS and RENO~\cite{Atif:2020glb}, which benefits from the fact that both experiments are located at the same reactor, while covering different baselines.
They report a 95\% CL exclusion in the region of $0.1\,\text{eV}^{2} \leq m^{2}_{41} \leq 7\,\text{eV}^{2}$.

\subsection{Rate~Approach}
Following the absolute rate approach, and~profiting from a compact and highly enriched $^{235}\text{U}$ core (HEU), the~STEREO collaboration was able to present the currently most precise absolute flux measurement in 2020~\cite{STEREO:2020fvd}.
The measurement is based on STEREO's phase-II data of 119 days of reactor-on and 211 days of reactor-off data combining all six cells.
STEREO reports a rate deficit of $(5.2\pm0.8~\text{[stat]}\pm2.3~\text{[sys]}\pm2.3~\text{[model]})\%$ compared to their model.
The deficit is consistent with the world average of previous measurements conduced by various experiments in the 1990s~\cite{Gariazzo:2017fdh,Giunti:2019qlt}. 
Previously, the~flux measurements at lowly enriched $^{235}\text{U}$ cores (LEU) were carried out by Daya Bay, Double Chooz, and~RENO at baselines of about 1\,km~\cite{DayaBay:2018heb,DoubleChooz:2019qbj,RENO:2018pwo}.
From these data, analyses of the dependency of the deficit on the reactor core composition over time have indicated that $^{235}\text{U}$ may be the main isotope driving the RAA~\cite{Giunti:2019qlt}.
A comparison between LEU and HEU experiments allows an additional handle on the isotopic source behind the deficit.
One can look at the common deficit of the $^{235}\text{U}$ component of the LEU experiments Daya Bay and RENO by conducting a fit, where the isotopic IBD yields of $^{235}\text{U}$ and $^{239}\text{Pu}$ are free, while those of $^{238}\text{U}$ and $^{241}\text{Pu}$ are constrained to the prediction~\cite{Giunti:2019qlt}.
Comparing this LEU result with the HEU result of STEREO yields a tension of 2.1 standard deviations~\cite{STEREO:2020fvd}. This indicates that $^{235}\text{U}$ might not be the sole source of the~deficit.

\section{Measurements of the Spectral~Shape}
\label{sec:shape}
\begin{figure}[H]
    \includegraphics[width=\linewidth]{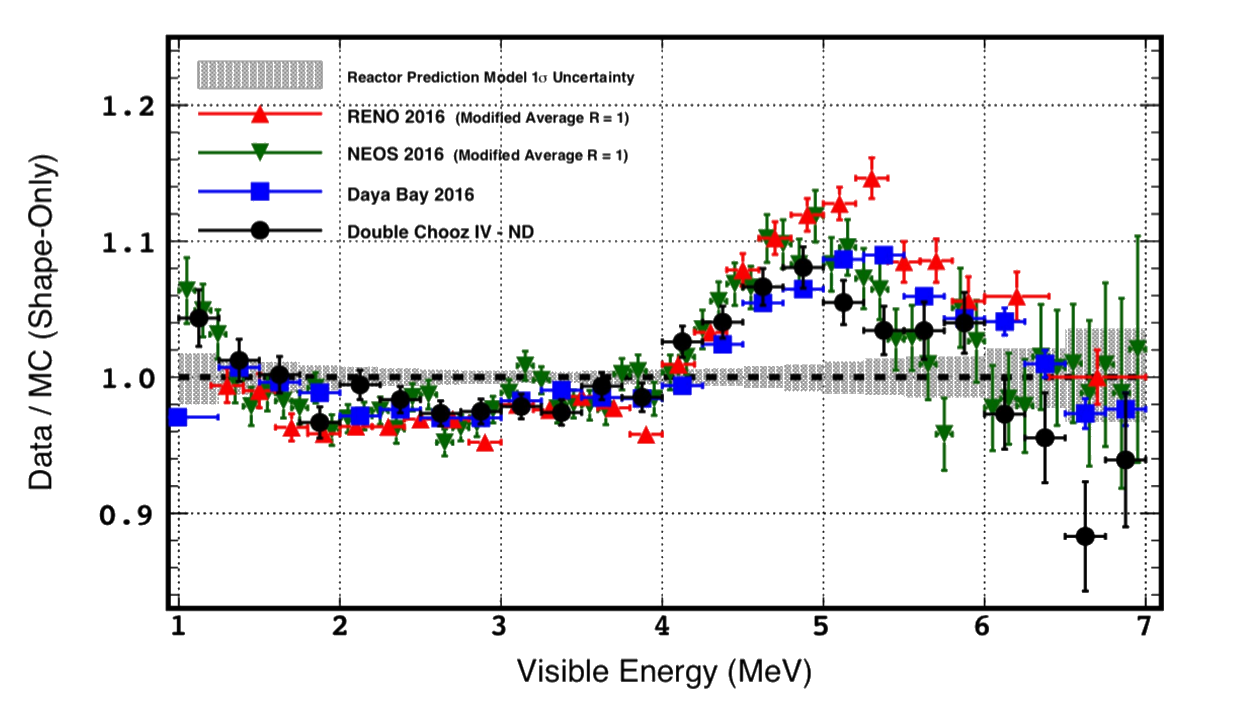}
    \caption{Comparison of spectral shape measurements of the LEU experiments. All results are normalised such that the average ratio equals unity. Reprinted from~\cite{DoubleChooz:2019qbj}, \copyright~2020.}
    \label{fig:shapeLEU}
\end{figure}

In 2014, distortions between the measured and predicted antineutrino energy spectrum, most prominently an excess around 5\,MeV, was reported and confirmed by the three kilometre-baseline reactor experiments: Double Chooz, Daya Bay and RENO~\cite{DoubleChooz:2014kuw,DayaBay:2015lja,RENO:2015ksa}.
To present date, such distortions are observed in even higher resolution as illustrated in Figure~\ref{fig:shapeLEU}. 
They are confirmed by additional experiments at short baselines.
The distortions correlate with reactor thermal power favouring reactor antineutrinos as cause of the distortion.
Earlier measurements by the CHOOZ and Goesgen collaborations show similar distortions, but~are not conclusive~\cite{Zacek:2018bij}.
As an exception, the~earlier measurement performed by the Bugey-3 collaboration exhibits a flat spectrum in disagreement with the findings above~\cite{Declais:1994su}.

Of particular interest is the comparison of spectra obtained from HEU and LEU reactors.
This allows disentangling of the contribution of individual isotopes to the overall spectral distortion~\cite{Buck:2015clx}.
\unskip

\subsection{LEU Measurements at~Kilometre-Baselines}
Daya Bay is providing a high statistics reactor antineutrino spectrum weighted by the IBD cross-section for model-independent predictions~\cite{resultDAYABAY}.
It shows 4.4-standard deviation local significance for an excess between 4 and 6\,MeV with respect to the commonly used Huber-prediction~\cite{conversion2}.
RENO reports a spectral shape mostly in agreement with Daya Bay's reference, also finding an excess around 5\,MeV corresponding to 3\% of their total number of IBD events~\cite{RENO:2016ujo}.
An even better agreement is found by Double Chooz~\cite{DoubleChooz:2019qbj}.
In their analysis, they additionally find a mild indication for a double-peak structure as origin behind the excess around 5\,MeV.

\subsection{LEU Measurements at Short~Baseline}
As the first short-baseline experiment, NEOS confirms the spectral distortion and excess around 5\,MeV with high significance~\cite{resultNEOS}.
Their findings are in good agreement with Daya Bay, as~can be seen in Figure~\ref{fig:shapeLEU}.
DANSS currently reports no significant distortions compared to the model predictions~\cite{resultDANSS}. However, ongoing improvements in systematics are likely to allow a refined quantitative statement in the~future.

\subsection{HEU~Measurements}
Apart from the LEU reactor experiments above, STEREO and PROSPECT also find a spectral distortion at HEU reactors.
STEREO uses an unfolding method to provide a pure $^{235}\text{U}$ spectrum in antineutrino energy~\cite{STEREO:2020hup}.
The collaboration finds an excess around 5.3\,MeV, rejecting the Huber prediction~\cite{conversion2} at 3.5 standard deviations.
A slightly better agreement is found with the updated spectrum of the summation method~\cite{Estienne:2019ujo}.
Moreover, the~PROSPECT collaboration finds excellent agreement of their unfolded spectrum compared to LEU results~\cite{PROSPECT:2020sxr}.
Their data prefer the measured LEU distortion at 2.2 standard deviations with respect to the model.
PROSPECT is able to disfavour $^{235}\text{U}$ as sole source of the spectral distortions seen at LEU reactors at 2.4-standard deviation confidence~level.

\subsection{Combined HEU and LEU-HEU~Analyses}
\begin{figure}[H]
    \includegraphics[width=\linewidth]{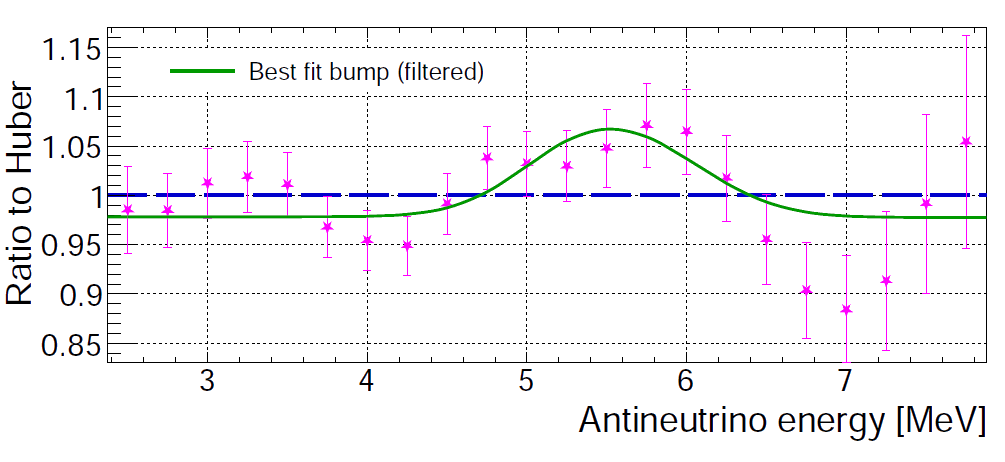}
    \caption{The jointly unfolded $^{235}\text{U}$ spectrum of the HEU experiments STEREO and PROSPECT relative to Huber's prediction. The~excess around 5\,MeV is modelled by a Gaussian. Reprinted from~\cite{ProspectStereo:2021lbs} under CC BY~4.0.}
    \label{fig:stereoprospect}
\end{figure}

As for the sterile oscillation analysis, collaborations are moving towards joint efforts exploiting detailed systematic treatments and mitigating site-dependent or detector-dependent effects.
Finding their spectra in agreement, the~HEU experiments PROSPECT and STEREO perform a joint unfolding process yielding the spectrum depicted in Figure~\ref{fig:stereoprospect}.
They report an excess around 5 to 6\,MeV at 2.4 standard deviations~\cite{ProspectStereo:2021lbs}.

A joint analysis between the LEU and HEU experiments has been carried out by PROSPECT and Daya Bay as well as NEOS and RENO.
PROSPECT and Daya Bay find their shape measurements to be consistent.
A combination reduces the degeneracy between the spectra of the $^{235}\text{U}$ and $^{239}\text{Pu}$ isotopes and allows reduction in the uncertainty of the $^{235}\text{U}$ spectral shape to 3\%~\cite{An:2021tyg}.
NEOS and RENO find comparable results to Daya Bay in their joint analysis~\cite{Atif:2020glb}.
They merely find small spectral features contained within their respective~uncertainties.

\section{Summary and~Outlook}
\label{sec:outlook}
Measurements at the LEU and HEU reactors give an overall coherent picture on the spectral shape distortion and start to allow identification of the contributions of individual isotopes to the overall shape anomaly.
At the same time, several refinements and revisions of the flux prediction are ongoing and offer possible explanations to the deviations from the prediction.
At the same time, a~large parameter space for sterile neutrino oscillations has been excluded by most collaborations with the exception of an indication for such oscillations by Neutrino-4.
There is still some relevant parameter space of the RAA to be explored before a definitive statement on sterile neutrinos can be drawn (cf.~Figure~\ref{fig:oscall}).
Several collaborations are working to reach this goal.
While data collection has been completed, the~majority of the current reactor experiments are still analysing and extending their full dataset or planning on follow-up experiments to gain higher sensitivity and significance.
DANSS plans to upgrade their detector with new scintillator strips giving higher light collection efficiency and better homogeneity~\cite{Svirida:2020zpk}. 
STEREO are currently analysing their phase-III data, which adds comparable statistics to their phase-II data~\cite{STEREO:2019ztb}.
Neutrino-4 is planning to establish a second experimental site which will accommodate a three-fold more sensitive detector in order to confirm their results~\cite{Serebrov:2020kmd}.
PROSPECT plans a detector upgrade possibly also involving deployment at a HEU and LEU reactor to resolve isotopic dependencies~\cite{Andriamirado:2021qjc}.
In addition to the reactor experiments, KATRIN has recently achieved exclusion limits at very high mass splittings~\cite{KATRIN:2020dpx}.
They could provide an additional competitive exclusion in the higher mass slitting range of reactor experiments in the future, as~indicated in Figure~\ref{fig:oscall}.
In addition, a~joint analysis between Daya Bay, STEREO, and~PROSPECT is investigated and would merge the joint works of each~collaboration.

\vspace{6pt} 




\funding{The author thanks for the support received by the Alexander von Humboldt-Stiftung.}

\institutionalreview{Not applicable}

\informedconsent{Not applicable}


\acknowledgments{The author thanks Prof. Dr. Olga Mena Requejo and Dr. Pablo Fern\'{a}ndez de Salas for their invitation to write this review article.}

\conflictsofinterest{The author declares no conflict of interest. The~funders had no role in the writing of the manuscript or in the decision to publish~it.} 



\end{paracol}
\reftitle{References}


\externalbibliography{yes}
\end{document}